\documentclass[prd,aps,11pt,
nofootinbib,tightenlines,superscriptaddress,preprintnumbers]{revtex4}

\preprint{BI-TP 2006/05}
\preprint{RBRC 597}

\def\lsim{\raise0.3ex\hbox{$<$\kern-0.75em\raise-1.1ex\hbox{$\sim$}}}
\def\gsim{\raise0.3ex\hbox{$>$\kern-0.75em\raise-1.1ex\hbox{$\sim$}}}

\usepackage{graphicx}

\newcommand{\beq}{\begin{equation}}
\newcommand{\eeq}{\end{equation}}
\newcommand{\bqa}{\begin{eqnarray}}
\newcommand{\eqa}{\end{eqnarray}}

\begin{document}

\author{Rudolf Baier}
\author{Paul Romatschke}
\affiliation{Fakult\"at f\"ur Physik, Universit\"at Bielefeld, 
D-33501 Bielefeld, Germany}
\author{Urs Achim Wiedemann}
\affiliation{Department of Physics and Astronomy, University of Stony
Brook, NY 11794, USA}
\affiliation{RIKEN-BNL Research Center, Brookhaven National Laboratory, Upton,
NY 11973-5000, USA}
\date{\today}

\title { Transverse flow in relativistic viscous hydrodynamics}

\date{\today}

\begin{abstract}
Hydrodynamic model simulations of Au-Au collisions at RHIC have indicated
recently, that with improved simulations in the coming years, it may be feasible 
to quantify the viscosity of the matter produced in heavy ion collisions. To this
end, a consistent fluid dynamic description of viscous effects is clearly needed.
In this note, we observe that a recently used, approximate form of the 2nd order
Israel-Stewart  viscous hydrodynamic equations of motion cannot account 
consistently for the transverse flow fields, which are expected to develop in heavy
ion collisions. We identify the appropriate equations of motion.
\end{abstract}

\maketitle

\section{Introduction}

Ultra-relativistic nucleus-nucleus collisions exhibit large collective phenomena, 
such as transverse flow. In the discussion of the dynamical origin of
this collectivity, relativistic hydrodynamics plays a central role. This is so, since 
the densities attained in nucleus-nucleus collisions imply very small mean free
paths ($< 1$ fm), a condition, which hydrodynamics can naturally account for, while
other approaches to multi-particle dynamics (such as parton cascades) have
difficulties to control. Indeed, simulations based on ideal fluid dynamics successfully
describe the main characteristics of soft momentum distributions in nucleus-nucleus
collisions at RHIC~\cite{QGP3,Heinz05,Heinz2}. 
However, the results of these simulations may depend significantly
on the modeling of the initial conditions and freeze-out. This raises the question
whether the success of ideal hydrodynamics is indicative of an almost
viscosity-free matter produced at RHIC, or to what extent matter of non-negligible
viscosity can be accounted for by ideal hydrodynamics due to a compensatory
choice of initial conditions and freeze-out~\cite{Baier}. To address this point, the study of the 
hydrodynamical equations of motion in the presence of viscous corrections is 
clearly needed.

Including dissipative corrections of ideal fluid dynamics to first order in a gradient
expansion is known to lead to an acausal dynamics, exhibiting significant unphysical
features~\cite{Hisc}. A consistent relativistic dissipative hydrodynamics requires a gradient
expansion to second order~\cite{I,IS79}. The application of this Israel-Stewart
theory to relativistic heavy ion collisions is at the very 
beginning~\cite{Muronga,MR04,Baier,Raju2,CH05,HSC04}. 
In this note, we show that a specific
approximation of the Israel-Stewart equations of motion~\cite{CH05,HSC04}, 
used recently, does not allow for a consistent description of transverse flow, 
and we identify the full equations of motion, which should be used.

For simplicity, we consider hot matter with an ultra-relativistic equation of state (EoS) 
$\epsilon =3 p$. In the absence of conserved charges (vanishing chemical
potential) and for finite shear viscosity $\eta$, the energy-momentum tensor takes
the form
\beq
T^{\mu \nu} = (\epsilon + p) u^{\mu} u^{\nu} - p g^{\mu \nu}
 + \Pi^{\mu \nu} \, .
\label{e1}
\eeq
Here $\epsilon,p$ are the energy density and pressure,  $u^{\mu}$ denotes the 
local fluid velocity, $u^{\mu} u_{\mu} = 1$, and $\Pi^{\mu\nu}$ is the traceless
shear tensor. The tensor $T^{\mu \nu}$ is conserved,
\beq
   d_{\mu} T^{\mu \nu} = 0 \, ,
\label{e3}
\eeq
where $d^{\mu}$ is the covariant derivative. In the Israel-Stewart theory,
the evolution equation of the shear
tensor is given by \cite{IS79,Baier}
\beq
\tau_{\Pi} \Delta^\mu_\alpha \Delta^\nu_\beta D \Pi^{\alpha \beta}
+\Pi^{\mu \nu}=\eta  \langle\nabla^\mu u^\nu\rangle 
\label{fulleq}
\eeq
with
$\nabla^\mu \equiv \Delta^{\mu \nu} d_\nu$,  $\Delta^{\mu\nu} \equiv  
g^{\mu \nu}-u^\mu u^\nu $, 
$\langle \nabla_\mu u_\nu\rangle \equiv \nabla_\mu u_\nu+\nabla_\nu u_\mu
-\frac{2}{3} \Delta_{\mu \nu} \nabla_\alpha u^\alpha$, 
and $D=u^{\mu}d_{\mu}$. In the limit of
vanishing relaxation time $\tau_{\Pi}$, one finds the defining equation
for $\Pi^{\mu\nu}$ in first order dissipative hydrodynamics. 
Here and in what follows, we neglect for simplicity other dissipative 
corrections (bulk viscosity, heat conductivity), as well as contributions from the
vorticity to (\ref{fulleq}). 

Eq.~(\ref{fulleq}) can be derived from kinetic theory, which is one way of determining
all possible terms entering the equation of motion~\cite{Baier}. 
An alternative 'short-cut' derivation is based
on the constraint that up to second order in a gradient expansion, the comoving entropy 
density $s^{\mu}$ cannot decrease \cite{I,Rischke:1998fq}
\bqa
T d_{\mu} s^{\mu} &=& \Pi_{\mu \nu}
[ - \tau_{\Pi} D \Pi^{\mu \nu}
+ \eta <\nabla^\mu u^\nu>]/(2 \eta) \nonumber \\
&\equiv& \frac{1}{2 \eta} \Pi_{\mu \nu}\Pi^{\mu \nu} \geq 0\, .
\label{e5}
\eqa
Motivated by this constraint, one uses in the recent literature~\cite{CH05,HSC04} the
equation of motion
\beq
\tau_{\Pi}  D \Pi^{\mu \nu}
+\Pi^{\mu \nu} = \eta <\nabla^\mu u^\nu> \, .
\label{e4}
\eeq
However, since $u_{\mu} \Pi^{\mu\nu} =0$, Eq.~(\ref{e5}) 
does not specify the components parallel to $u_\mu$ in (\ref{e4}). Thus, 
Eq.~(\ref{e4}) is based on (\ref{e5}) and the additional approximation that
$u_{\mu} D \Pi^{\mu \nu} =0$, which one obtains by contracting (\ref{e4}) with $u_\mu$.
On the other hand, contracting the full Israel-Stewart equation (\ref{fulleq}) with $u_\mu$, one
finds $u_{\mu} D \Pi^{\mu\nu} = -  \Pi^{\mu\nu} D u_\mu$, which is not an additional
constraint but a simple consequence of the orthogonality $u_{\mu} \Pi^{\mu\nu} =0$. 
It follows that (\ref{e4}) involves the additional assumption $\Pi^{\mu\nu} D\, u_{\mu} = 0$.
At least for the physically relevant case of a radially symmetric collision region, this latter
constraint implies necessarily (see the appendix for details of this argument) 
\beq
 D u^{\mu} = 0. 
\label{e6}
\eeq
We emphasize that (\ref{e6}) is a necessary consistency condition for the
approximate equation of motion (\ref{e4}),
but it is in general not satisfied for the full Israel-Stewart equation of motion (\ref{fulleq}). 

\section{Solution for ideal hydrodynamic evolution}

Is the additional assumption (\ref{e6}) acceptable for the simulation of
relativistic heavy ion collisions? To address this question, we study here the
resulting solution of ideal hydrodynamics, which serves as the reference in 
comparison with viscous effects~\cite{CH05,HSC04,Teaney03,Teaney04}. Our 
starting point is Eq.~(\ref{e6}), which in the sense of a 'proof by contradiction'
is assumed to be valid in this section. We also use 
\begin{eqnarray}
\nabla^\mu p = 0  \, ,
\label{e8}\\
D\epsilon = -(\epsilon + p)\nabla_\mu u^{\mu}
\, \longrightarrow \, 
Dp = -\frac{4}{3} p \,  \nabla_\mu u^{\mu}  \, ,
\label{e9}
\end{eqnarray}
which follow from the conservation laws (\ref{e3}), together with (\ref{e6}). 
We consider a system with longitudinal boost-invariance
\cite{Bjorken} and cylindrical symmetry in the transverse direction.
The independent variables are $\tau,r,\phi$ and rapidity $\eta$.
Because of the symmetries, the pressure $p = p(\tau,r)$ and radial
flow velocity $v= v(\tau,r)$,
\beq
v = \frac{u^r}{u^{\tau}} \, \,  , v \le 1 \, ,
\label{e10}
\eeq 
are functions of $\tau$ and $r$ only.
Using the metric components,
$g_{\tau \tau} = 1, g_{r r}= -1, g_{\eta \eta} = - \tau^2,
g_{\phi \phi}= - r^2$,
the relativistic hydrodynamic equations become
\bqa
( \partial_r + v \partial_{\tau}) \,  p(\tau,r) &=& 0 \, , \label{e11a} \\ 
( \partial_{\tau} + v \partial_r) \, p(\tau,r) +
\frac{4}{3} p(\tau,r) [\gamma^2 ( \partial_r + v \partial_{\tau})
\, v(\tau,r)  + \frac{1}{\tau} + \frac{v}{r} ] &=& 0 \, , \label{e11b} \\
( \partial_{\tau} + v \partial_r) \, v(\tau,r) &=& 0 \, ,
\label{e11}
\eqa
where $\gamma^{-2} = 1 - v^2$, and
 $\partial_r = \frac{\partial}{\partial r}$,
$\partial_{\tau} = \frac{\partial}{\partial {\tau}}$.
This set of quasilinear partial differential equations can be solved
by the method of characteristics. 

From Eq.~(\ref{e11}), one finds that along the characteristic $\tau,r(\tau)$ that fulfills
the equation $\frac{dr}{d\tau}=v$, the value of $v$ is constant,
\beq
\frac{dv}{d\tau}=0.
\eeq
Therefore, one can integrate $\int dr= \int v d\tau$, obtaining
\beq
v\tau-r={\rm const}=f(v),
\label{se11}
\eeq
where we used $\frac{dv}{d\tau}=0$ to rewrite the constant as an
arbitrary function of $v$. Eq.~(\ref{se11}) is the general solution
of Eq.~(\ref{e11}). It has been discussed previously 
in a different context, see e.g. Biro \cite{Biro:2000nj,Biro:1999eh}.

We now use
\beq
\gamma^2(\partial_r+v\partial_\tau)v=\partial_r v,\qquad
p=\tilde{p}\, (r\tau)^{-4/3}\, ,
\eeq
to rewrite Eq.~(\ref{e11b}) as
\beq
(\partial_\tau+v\partial_r)\tilde p = - \frac{4}{3} \tilde p\, \partial_r v.
\eeq
Using again the method of characteristics, we find
\beq
d\, \ln{\tilde{p}^{3/4}} = - (\partial_r v)\, d\tau,\qquad v\, d\tau = dr.
\label{pse11b}
\eeq
Since the second equation is the same characteristic as before we know
that along this characteristic $\frac{dv}{d\tau}=0$. Rewriting
\beq
\partial_r\, v=\frac{1}{\tau-f^\prime(v)}
\label{also}
\eeq
from Eq.~(\ref{se11}) with $f^\prime(v)=\frac{df}{dv}$, 
one can thus integrate Eq.~(\ref{pse11b}), obtaining
\beq
\tilde{p}^{-3/4}=(\tau-f^\prime(v))\, \times {\rm const.} =
(\tau-f^\prime(v)) g(v),
\eeq
where we introduced another arbitrary function $g(v)$ for the same
reasons as above. In summary, we find for the general solution of Eq.~(\ref{e11b})
\beq
p^{3/4}=(r\tau)^{-1}(\partial_r v) \bar{g}(v)\, ,
\label{se11b}
\eeq
where $\bar{g}(v)=g^{-1}(v)$. 
The remaining Eq.~(\ref{e11a}) can be used to constrain the
functions $f(v)$ and $g(v)$. Inserting Eq.~(\ref{se11b}) into
Eq.~(\ref{e11a}) and using Eqs.~(\ref{e11},\ref{se11},\ref{also}) we find
\bqa
 - v\, f\, g\, (f^\prime)^2&&\nonumber\\
+\tau\left[(1-v^2)f\, f^\prime\, g^\prime+g\, (1+v^2) 
(f^\prime)^2+g f(3vf^\prime+(1-v^2)
f^{\prime\prime})\right]&&\nonumber\\
-\tau^2\left[(1-v^2)(f+vf^\prime)g^\prime+g\left(2vf+(2+3v^2)
f^\prime+v(1-v^2)f^{\prime\prime}\right)\right]&&\nonumber\\
+\tau^3\left[(1+2v^2)g+v(1-v^2)g^\prime\right]&=&0.
\label{consteq}
\eqa
For solutions relevant in heavy ion collisions, we require that at all spatial points $r$ 
and for all times $\tau >0$,  the pressure and its time derivative are finite. 
In particular,  we require  $0<p(\tau,r=0)<\infty$, and 
$\vert \left.\partial_\tau p(\tau,r =0) \right. \vert <\infty$.
According to Eq.~(\ref{se11b}), the first condition leads either to 
(i) $\lim_{r\ll1} \partial_r v\sim r $ or to (ii) $\lim_{r\ll1}\bar{g}(v)\sim r$. 
In the next paragraph, we shall show that case (i) is realized. 
Because of Eq.~(\ref{e11}),  case (i)  implies $\partial_\tau v(\tau,r=0)=0$. 
For case (ii), one considers  $\partial_r v|_{r=0}\neq0$. Then, to satisfy $\partial_\tau p<\infty$, 
one has to require $\lim_{r\ll1}\partial_\tau \bar{g}(v)\sim r$,
and this leads again to $\partial_{\tau}v(\tau,r=0)=0$.
As a consequence of Eq.~(\ref{se11}) we thus find in both cases $v(\tau,r=0)=0$.

Since $v(\tau,r=0)=0$, obviously $v$ does not depend on $\tau$ at $r=0$. Thus, for
Eq.~(\ref{consteq}) to hold at all times $\tau$, each power of $\tau$ in (\ref{consteq})
has to vanish separately. One can exploit this property in a neighborhood around
$v=0$ by doing a Taylor expansion of (\ref{consteq}) in $v$ around $v=0$. 
One then finds that not only the $v$-dependent prefactors of the powers $\tau^n$
in (\ref{consteq}) must vanish at $v=0$, but also all their derivatives must vanish. 
Thus, these prefactors must be zero and one finds
\beq
f(v)=0,\qquad g(v)= {\rm const} \times   \frac{1}{v}(1-v^2)^{3/2}\, ,
\label{final2}
\eeq
which leads to
\beq
v (\tau,r) =\frac{r}{\tau},
\qquad p (\tau,r) = {\rm const}  \times  (\tau^2-r^2)^{-2}\, .
\label{final}
\eeq
This velocity  field $v$ and pressure $p$ are physical for $r<\tau$, only.
At $r=0$ the pressure behaves as
$p(\tau, r=0) \simeq 1/\tau^4$, which differs significantly from the
Bjorken cooling law  $p(\tau) \simeq \tau^{-4/3}$ \cite{Bjorken}.
This exact analytic solution (\ref{final}) of  ideal
relativistic hydrodynamics in $(1+3)$ dimensions,
constrained by $Du^{\mu} = 0$ and  restricted by 
 boost invariance and cylindrical symmetry,
has also been discussed previously in \cite{Csorgo:2003rt}.

For hydrodynamic simulations of central heavy ion collisions, the initial conditions
satisfy usually $v(\tau_0, r) = 0$ for finite time $\tau_0$ and all values of $r$ 
\cite{CH05,HSC04,Teaney04}. This initial condition is incompatible with the
requirement (\ref{final2}) that $f(v) = 0$ in a finite region around $v=0$, and so 
it is incompatible with Eq.~(\ref{consteq}). Moreover, in numerical solutions
of the full ideal hydrodynamic equations of motion (for which $D u^\mu \not= 0$
in general), one finds that the velocity field for small but fixed values of $r$ increases
with $\tau$ for some time (see e.g. Fig.~1 in \cite{Teaney04}). In contrast, for the
solution (\ref{final}), the velocity $v(\tau,r={\rm const})$ decreases with $\tau$.  
This illustrates, that $Du^{\mu} = 0$ is not a physically justified
approximation in the context of heavy ion collisions, and so it should not be adopted for the
refined formulation of viscous hydrodynamics. For reliable results, 
the full Israel-Stewart equation (\ref{fulleq}) rather than Eq.~(\ref{e4})
has to be used.

\vspace{2cm}

\acknowledgments

PR was supported by BMBF 06BI102.
UAW acknowledges the support of the Department of Physics and Astronomy,
University of Stony Brook, and of RIKEN, Brookhaven National Laboratory, and the 
U.S. Department of Energy [DE-AC02-98CH10886] for providing the facilities essential 
for the completion of this work. 

\appendix
\section{Equation (\ref{e4}) implies $D u^\mu=0$}

In this appendix, we consider a longitudinally boost-invariant system
with cylindrical symmetry in the transverse plane. We show that approximating
the full hydrodynamic equations (\ref{fulleq})  by (\ref{e4})  amounts to the
assumption  $D u^\mu=0$. We start from the observation that Eq.~(\ref{fulleq}) 
agrees with Eq.~(\ref{e4}), if
\beq
\Pi^{\mu \alpha} D u_\alpha=0.
\eeq
Since only $u^\tau,u^r$ are non-vanishing, we find (choosing $\mu=r$)
\beq
\Pi^{r \tau} D u^\tau - \Pi^{r r} D u^r =0.
\eeq
Using $u_\nu \Pi^{\nu \mu} =0$ and $u^\tau=\sqrt{1+(u^r)^2}$, this expression
can be written as
\beq
\Pi^{r r} \left( \left(\frac{u^r}{u^\tau}\right)^2 - 1\right) D u^r =0.
\eeq
Since $u^r$ is in general unequal $u^\tau$, this means that $D u^\mu=0$ if
$\Pi^{\mu \nu}$ is non-vanishing. In contrast, for the full viscous equation of
motion (\ref{fulleq}), $D u^\mu \not= 0$ in general.

\end{document}